\documentclass[twocolumn,a4paper,10pt]{article}
\usepackage{amsmath,amsfonts}
\usepackage{graphicx}
\usepackage{epsfig}
\usepackage{float}
\usepackage{subfigure}
\usepackage{times}
\usepackage{epsfig}
\usepackage[small,hang]{caption2}
\makeatletter

\newcounter{numbersec}
\renewcommand{\section}[1]{\par\noindent\stepcounter{numbersec}
\par
\vspace{6pt}
\noindent\textbf{\large   \arabic{numbersec} \hspace*{0.3cm} #1 }
\par
\vspace{2pt}
}
\renewcommand{\subsection}[1]{
\par
\vspace{6pt}
\noindent\textbf{#1}
\par
}
\renewcommand{\subsubsection}[1]{%
\par
\vspace{6pt}
\textbf{#1.}
}

\newcommand{\Abstract}{\par\vspace{6pt}\noindent\textbf{\large Abstract}\par\vspace{2pt}}
\newcommand{\Acknowledgments}{\par\vspace{6pt}\noindent\textbf{\large Acknowledgments }\par\vspace{2pt}}

\newenvironment{References}{
\par\vspace{6pt}\noindent\textbf{\large References}\par\vspace{2pt}
\begin{small}\begin{list}{ }{
\itemsep0mm \parsep0mm\labelsep0mm\leftmargin0mm
}}
{\end{list}\end{small}}

\usepackage[normalem]{ulem}
\usepackage{color}

\makeatother

\title{\vspace*{-12mm}
\LARGE \sc \textbf{  
Revisiting the amplitude modulation in wall-bounded turbulence:\\ towards a robust definition \\
}}
\author{ \Large \bf \textit{ 
Eda Dogan$^1$, Ramis {\"O}rl{\"u}$^1$, Davide Gatti$^2$,}
\\ \Large \bf \textit{Ricardo Vinuesa$^1$ and Philipp Schlatter$^1$}  \\ \\
\bf  $^{1}$ \textit{ Linn{\'e} FLOW Centre, KTH Mechanics, 
Stockholm, Sweden } \\
\bf  $^{2}$ \textit{ Institute of Fluid Mechanics, Karlsruhe Institute of Technology, 
Karlsruhe, Germany} \\\\
\underline{\bf eda.dogan@mech.kth.se}
}
\date{}
\begin{document}

\maketitle
\thispagestyle{empty}

%
%
\Abstract

The present study revisits the amplitude modulation phenomenon, specifically for the robustness in its quantification. To achieve this, a well-resolved large-eddy simulation (LES) data set at $Re_\theta~{\approx}~8200$ is used. First, the fluctuating streamwise velocity signal is decomposed into its small- and large-scale components using both Fourier filters and empirical mode decomposition (EMD), allowing the comparison among different separation filters. The effects of these filters on various definitions for quantifying the amplitude modulation have been discussed. False positive identification of the amplitude modulation has also been tested using a randomised signal. Finally, the impact of the inclination angle of the large-scale structures on the modulation quantification has been assessed.


%
%
\section{Introduction}

As large scales become more energetic and pronounced in high-Reynolds-number wall-bounded turbulent flows, their role throughout the boundary layer, especially on the small scales in the near-wall region, has been the focus of many recent investigations. One particular phenomenon that took attention by many is the amplitude modulation of the small-scale fluctuations in the near-wall region by the large scales in the outer region. Quantifying this amplitude modulation (AM) is not straightforward due to the non-linear nature of the phenomenon (Mathis \emph{et al.}, 2009), and therefore various approaches have been used in the literature. The first step into investigating the scale interactions is to decompose the fluctuating velocity signal into its large- and small-scale components. A common approach to achieve this is to use a cut-off spectral filter, typically applied in Fourier space. This  was done by Hutchins and Marusic (2007), Mathis \emph{et al.} (2009) and many others thereafter. The filter is chosen based on a cut-off wavelength (in space or time) that separates the inner and outer peaks in the pre-multiplied energy spectra of the streamwise velocity fluctuations, $u$. Depending on the available data set, filters can be defined as either temporal or spatial filters or a combination of both. Another approach to decompose the scales, which was used quite extensively by Leschziner and co-workers ({\it e.g.}\ Agostini and Leschziner, 2014
), is the Hilbert-Huang empirical mode decomposition (EMD) which assumes that the data consists of intrinsic modes of oscillations called intrinsic mode functions (\emph{imf}s).

To quantify the modulation, Mathis \emph{et al.} (2009) defined a correlation coefficient, $R$, between the large-scale velocity fluctuations, $u_{L}$, and the filtered envelope of the small-scale fluctuations, $E_{L}\left(u_S\right)$, 
\begin{equation} \label{eq:R}
R = \frac{\overline{{u_L^+}~{E_L\left(u_S^+\right)}}}{{\sigma_{u_L^+}}~{\sigma_{E_L\left(u_S^+\right)}}}~~~,
\end{equation}
where the overbar indicates the time-average operator, $\sigma$ is the standard deviation of the quantity in the subscript and $\left({\cdot}\right)^+$ represents inner scaling which is defined through the characteristic inner scales: the friction velocity $U_\tau$ (defined as $U_{\tau}=\sqrt{\tau_{w}/\rho}$, where $\tau_{w}$ is the mean wall-shear stress and $\rho$ is the fluid density) and the viscous length scale $\nu/U_{\tau}$, where $\nu$ is the kinematic viscosity.
The recent review paper by Dogan \emph{et al.}\ (2018) gives an overview of the different approaches to investigate the AM. They used two main aspects to categorize the studies in the literature on this topic: (i) the method for decomposing the scales, specifically Fourier filters and empirical mode decomposition (EMD), and (ii) how the modulation was quantified, namely single-point, two-point correlations and the scale-decomposed skewness term of the velocity fluctuations. They studied the various aspects on a single data set from a well-resolved large-eddy simulation (LES) by Eitel-Amor \emph{et al.} (2014). The present contribution will use the same data set to further explore the effects of secondary filtering in the $R$ definition by Mathis \emph{et al.} (2009), applied on the envelope of the small-scale fluctuations, on the quantification of the modulation. Additionally, the robustness of the definition will be investigated with random signals and the time shifts between the scale components.
\vspace{-3mm}
\section{Data set and spectral analysis} \label{sec:dataset}

The data set used in this study is the well-resolved LES of a spatially developing zero-pressure gradient turbulent boundary layer by Eitel-Amor \emph{et al.}\ (2014).  
For this data set, the domain starts at a low (laminar) $Re_\theta=180$, which is the momentum-thickness-based Reynolds number where $U_\infty$ is the free-stream velocity and is the momentum thickness. After tripping (Schlatter and \"Orl\"u, 2012), the flow is completely turbulent at around $Re_\theta~{\approx}~600$ and the maximum $Re_{\theta}$ shortly before the outflow is 8300. For the present contribution, the position $Re_\theta~{\approx}~8200$ is investigated since long time series of the flow are available for that $Re_\theta$. The reader is referred to Eitel-Amor \emph{et al.}\ (2014) for further details about the simulation, the numerical setup and its turbulence statistics.

\begin{figure*}[h!]
\centering
\subfigure[]{
\centering
\includegraphics[width=0.32\linewidth]{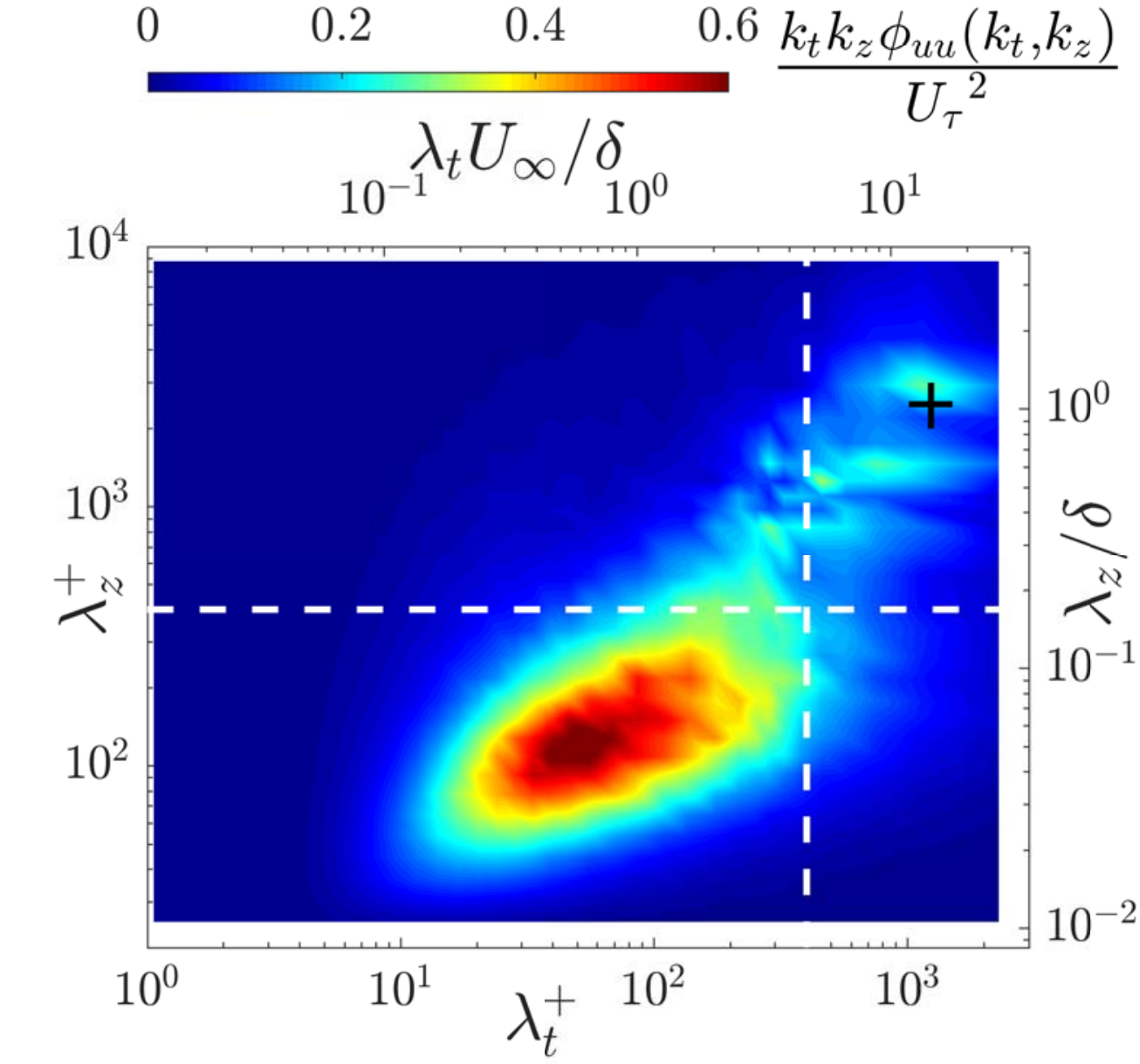} \label{fig:2Dspec1}}
\subfigure[]{
\centering
\includegraphics[width=0.32\linewidth]{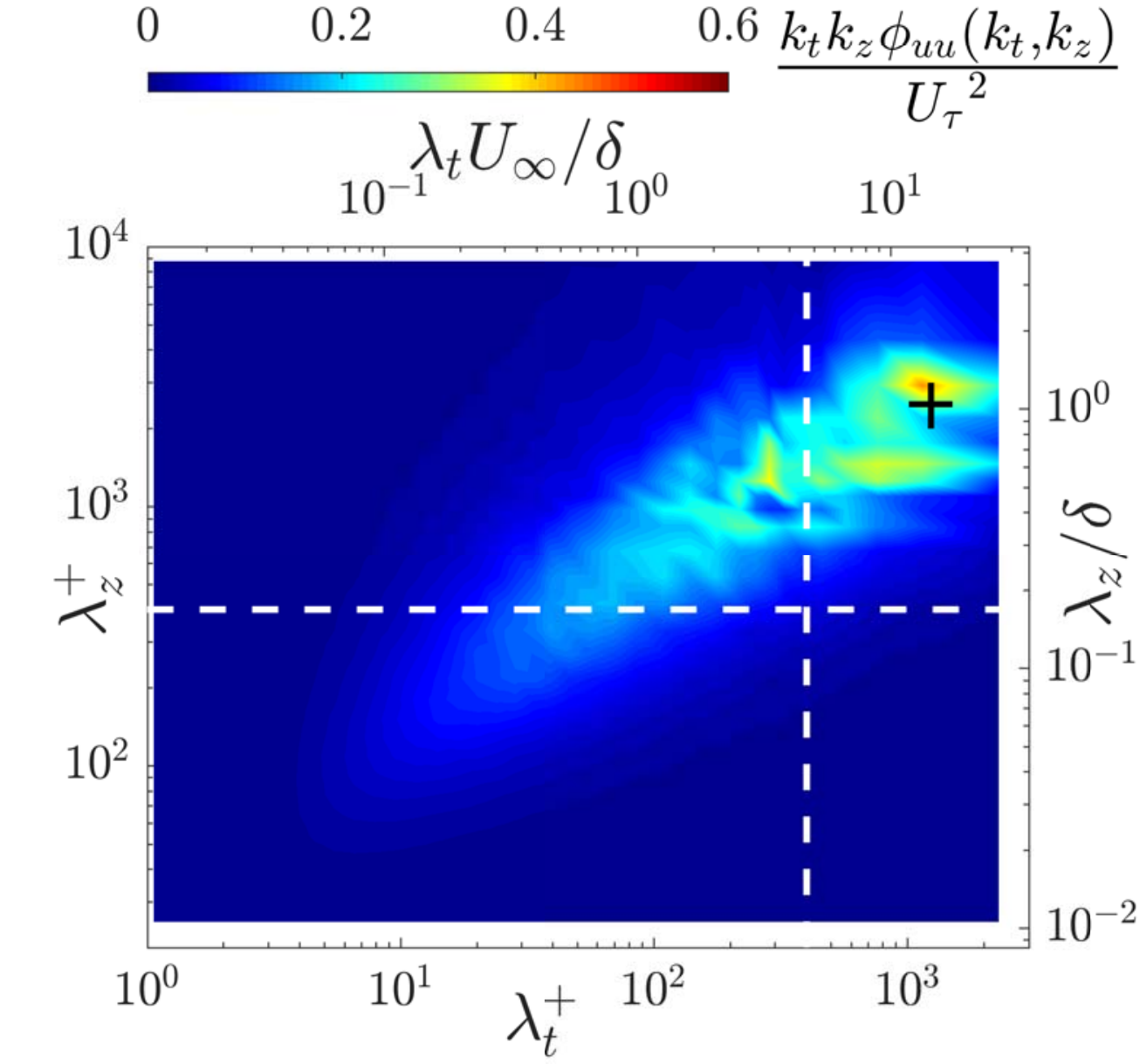} \label{fig:2Dspec2}}
\vspace{-2mm}
\caption{\label{fig:2Dspec} Contour maps of the inner-scaled pre-multiplied  2D energy spectra of the streamwise velocity fluctuations, $k_tk_z\phi_{uu}/U_{\tau}^{2}$, for $Re_\theta=8200$ at (a) $y^+{\approx}15$ and (b) $y^+{\approx}180$.\ The ordinates show spanwise wavelength, $\lambda_z$, in both inner (left) and outer (right) scaling. The abscissae show the wavelength in time, $\lambda_t$, in inner (bottom) and outer (top) scaling. Dashed white lines: the cut-off wavelengths for the spectral filter in both domains, i.e. $\lambda_t^+{\approx}400$ and $\lambda_z^+{\approx}400$. Black cross: the outer peak location, $\lambda_t{\approx}10\delta/U_\infty$ and $\lambda_z{\approx}\delta$.}
\end{figure*}

To investigate the distribution of energy across various scales in the boundary layer, contour maps of the energy spectra of the streamwise velocity fluctuations are extensively used, and since the current data set provides both spanwise and temporal information, 2D spectral maps are presented here. Figure \ref{fig:2Dspec} shows the inner-scaled pre-multiplied 2D spectral contour maps of the streamwise velocity fluctuations for $Re_\theta~{\approx}~8200$ at wall-normal locations of $y^+{\approx}15$ and $y^+{\approx}180$, which exhibits the two energy peaks in the boundary layer, {\it i.e.}\ near-wall and outer peak, respectively. The two panels at different wall-normal locations show the footprint of the large scales, {\it i.e.}\ the larger wavelengths in both directions located in the outer region (see Figure \ref{fig:2Dspec2}), on the region close to the wall (see Figure \ref{fig:2Dspec1}), confirming the superposition effect of the large scales on the near-wall region (Hutchins and Marusic, 2007). This spectral representation also allows assigning the energetic wavelengths of the streamwise fluctuating velocity signal to its large- and small-scale components, which will be further explored in the following section.
%
\subsection{Scale decomposition} \label{sec:scale}
In the present contribution we will handle the scale decomposition using two methods: spectral cut-off filters, {\it i.e.}\ Fourier filters, and EMD. Regarding the Fourier filters, the dashed lines representing the cut-off locations at $\lambda_t^+{\approx}400$ and $\lambda_z^+{\approx}400$ in Figure \ref{fig:2Dspec} reasonably demarcate the two peaks, {\it i.e.}\ the near-wall and outer peak. This is accepted as the sufficient criteria to set the Fourier filters to decompose the scales in the flow using spectral maps (note that different cut-off filters were also tested for the robustness of the filter choice and negligible differences were observed for the considered analyses, as previously suggested by Mathis \emph{et al.}\ (2009)). Implementing two cut-off filters in both directions (denoted 2DS filter here), we define the small scales as those with wavelengths smaller than the cut-off, both in time and span (with large scales being defined as the three remaining quadrants) giving them a robust representation.

For EMD, as previously mentioned, the underlying idea is that any data can be represented through different \emph{imf}s. 
Here, the extraction algorithm of these modes is adapted from the implementation of FABEMD (fast and adaptive bi-dimensional EMD, see Bhuiyan \emph{et al.}, 2008) by Cormier \emph{et al.}\ (2016), which was applied to a direct numerical simulation (DNS) of a channel flow at $Re_\tau=1000$. The FABEMD code is run to obtain 10 modes. The number of modes assigned to define small scales, $n_{SS}$, is in principle a free parameter that has an important impact on the separation of the scales. In our case, we found $n_{SS}=5$ to show optimal behaviour for the distribution of the variance profiles across the whole boundary layer, see Dogan \textit{et al.}\ (2018) for further discussion. 
The large scales comprise the residual of the signal once the small scales are assigned to their modes.

\begin{figure*}[h!]
\centering
\subfigure[]{
\centering
\includegraphics[width=0.32\linewidth]{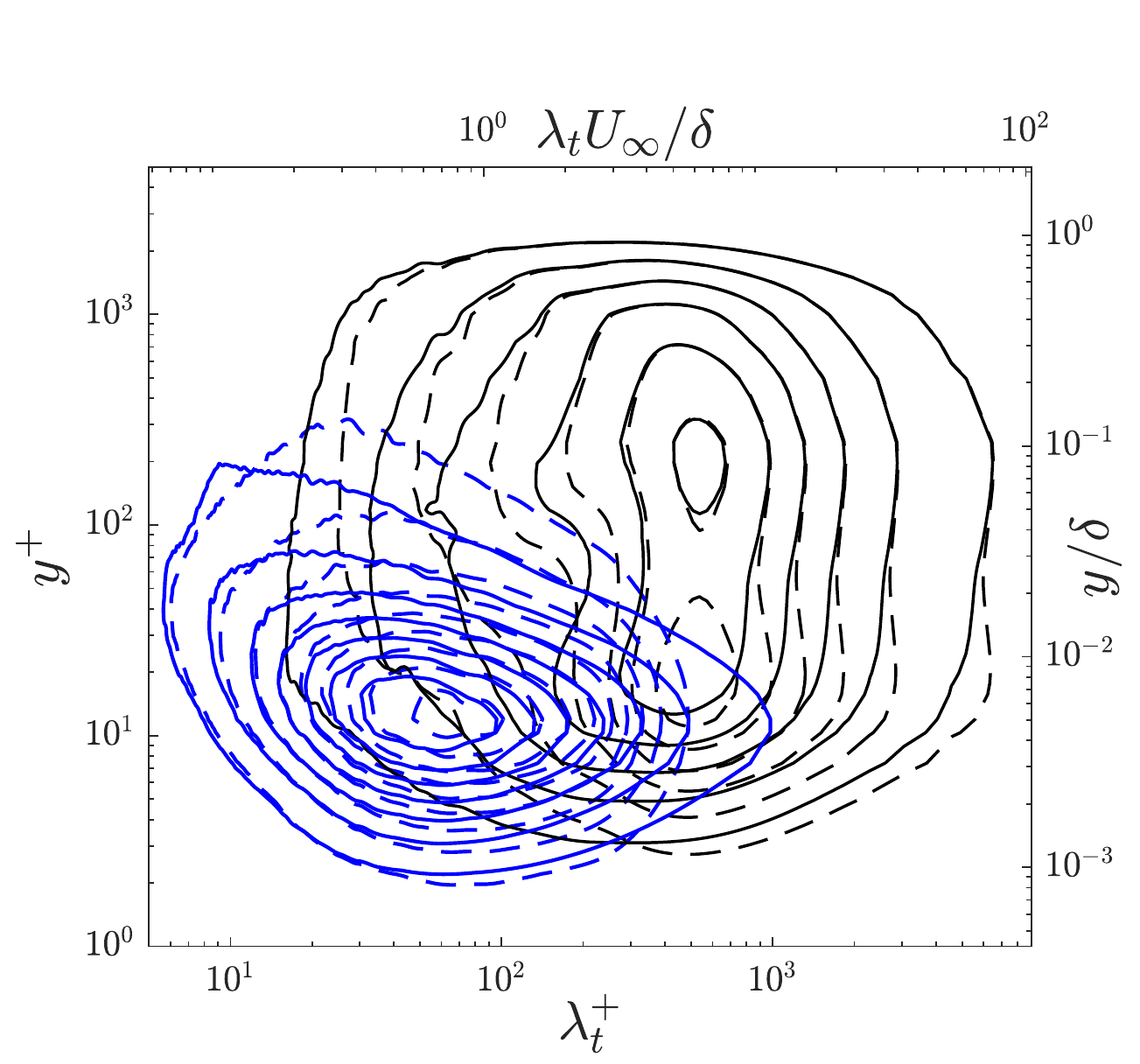} \label{fig:timespec}}
\subfigure[]{
\centering
\includegraphics[width=0.32\linewidth]{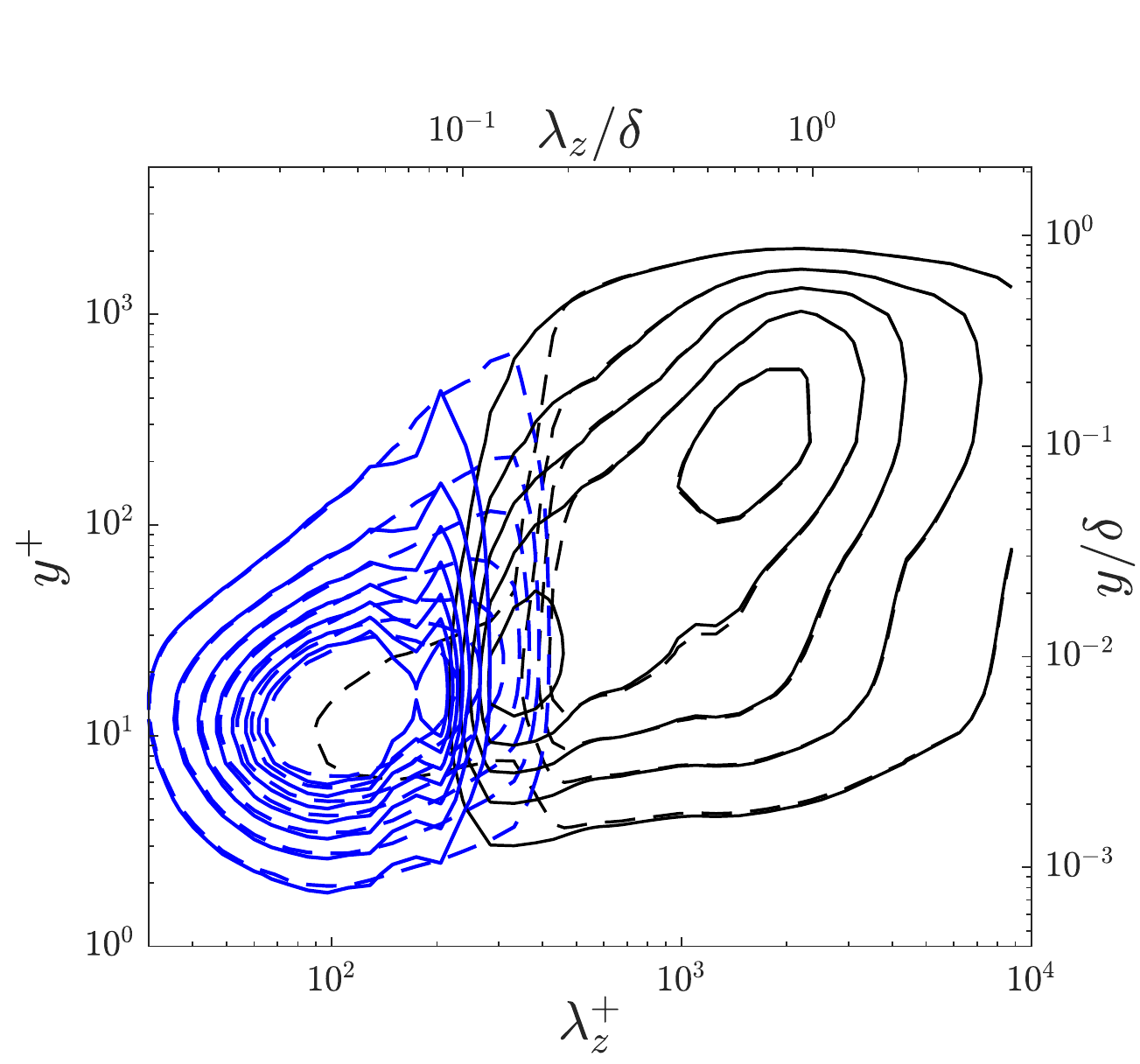} \label{fig:spanspec}}
\vspace{-2mm}
\caption{\label{fig:scalesspec} Contour plots of the inner-scaled pre-multiplied energy spectra of the streamwise velocity fluctuations for the decomposed scales at $Re_\theta=8200$. The energy levels for the small scales are represented by blue color and the large scales by black. The dashed contour lines represent the scales from 2DS Fourier filter, and the solid contour lines are from EMD. The abscissae show (a) time wavelength $\lambda_t$ and (b) spanwise wavelength $\lambda_z$, in  inner (bottom) and outer (top) units. The ordinates show the wall-normal location, $y$, in inner (left) and outer (right) units. The increment and the minimum contour level for the contours are (a) 0.2 and 0.2 (b) 0.3 and 0.3, respectively.}
\end{figure*}

A comparison between different decomposition methods can be assessed by analysing the energy spectra of the decomposed velocity signals. Figure \ref{fig:scalesspec} shows the inner-scaled pre-multiplied energy spectra of the streamwise velocity from small and large scales obtained through 2DS Fourier filter and EMD. The figure clearly indicates that the time spectra exhibits a significant overlap between the scales and makes it hard to define a meaningful cut-off to distinguish them; therefore it is advantageous to have both temporal and spatial information to define a 2D filter, at least for the present moderate $Re$ range. For higher $Re$, defining the cut-off through temporal period (or streamwise wavelength if Taylor's hypothesis is applied) has been a common approach, by Mathis \emph{et al.}\ (2009) and others, to decompose the scales, especially for hot-wire measurements where only time information is available. On the other hand, the spanwise spectrum shows a clear distinction between small and large scales. This demarcation is very sharp for the spectral filter, while a limited overlap for the EMD filter exists. It is quite remarkable that the difference of the energy levels of the scales obtained from the two decomposition methods is not significant. This shows the efficacy of both methods in adequately decomposing the scales.

\section{Results and Discussion}
To assess the influence of large scales in the outer region (at a height $y_1$) on the small scales in the near-wall region (at a height $y_2$), in principle simultaneous measurements at these two positions in the boundary layer are necessary. However, such a technique is not easy in experiments, which lead Mathis \emph{et al.}\ (2009) to define the correlation coefficient $R$, Equation (\ref{eq:R})  using  single-point hot-wire measurements although verified as a sufficient surrogate with two-point measurements. Numerical simulations, on the other hand, naturally provide simultaneous multi-point data (in fact whole velocity fields), and consequently Bernardini and Pirozzoli (2011) presented the correlation between the scales as 2D covariance maps from their DNS data. We use their quantification of the modulation, using the covariance, $C_{AM}$, {\it i.e.}\ the unnormalised form of the $R$ coefficient. In this definition, one aspect that did so far not receive sufficient attention is the effect of the filters involved in the calculation of each term, specifically, how the scales are decomposed, the augmented signal is calculated and  the envelope is filtered.
\subsection{Different filters}

\begin{figure*}[htbp!]
\centering
\subfigure[]{
\centering
\includegraphics[width=0.26\linewidth]{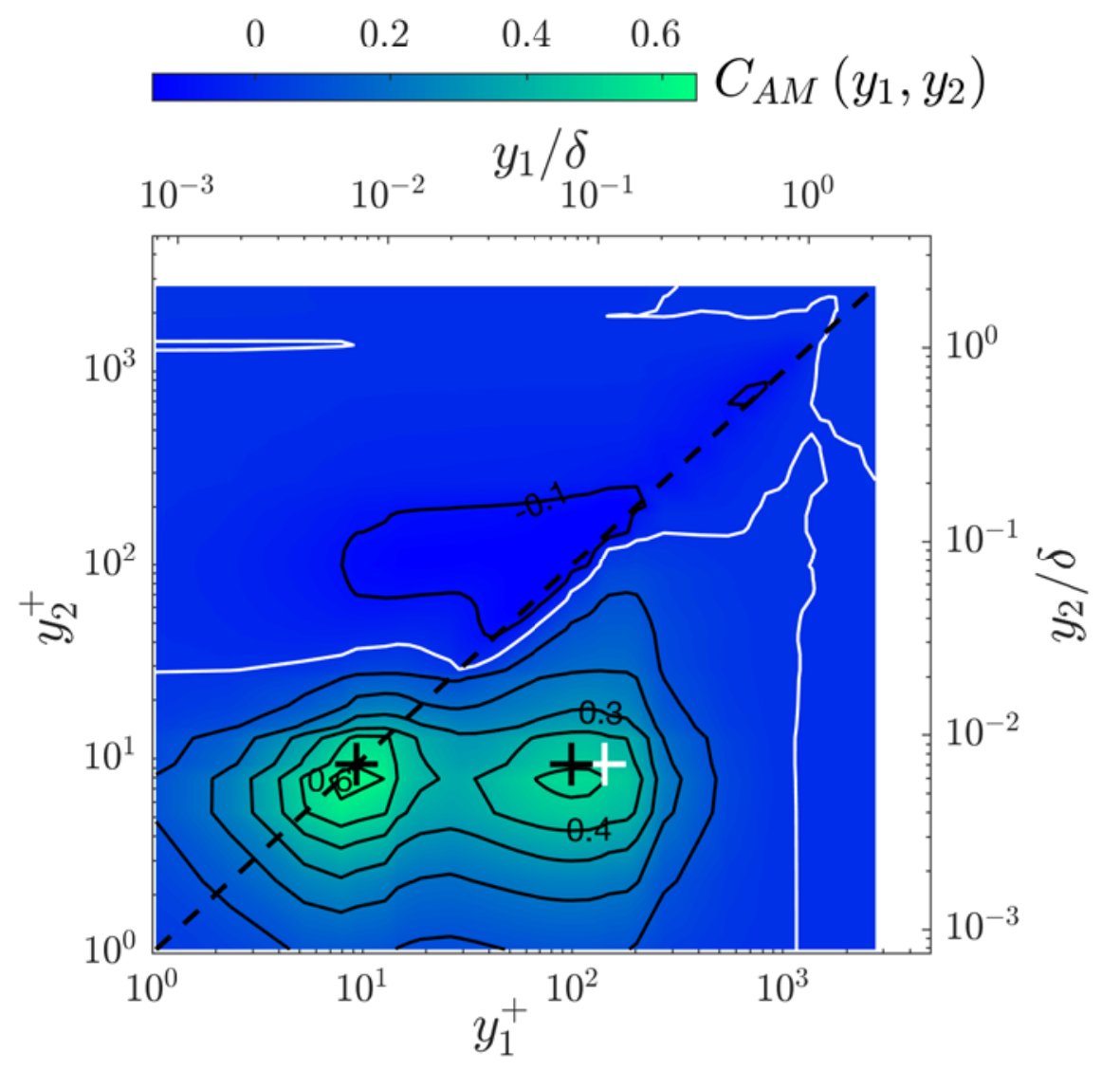} \label{fig:2pspan}}
\subfigure[]{
\centering
\includegraphics[width=0.26\linewidth]{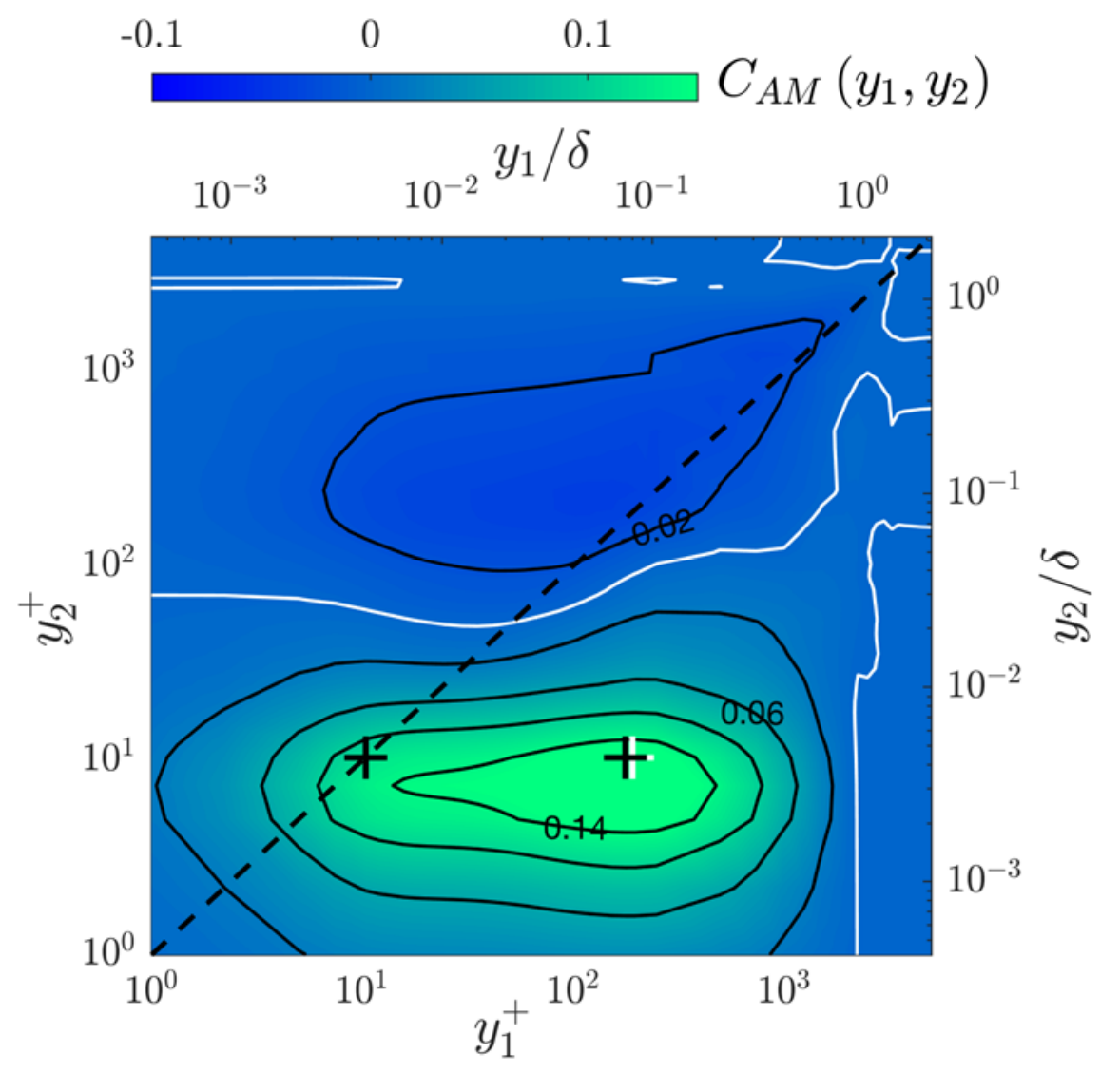} \label{fig:2pspantime}}
\subfigure[]{
\centering
\includegraphics[width=0.26\linewidth]{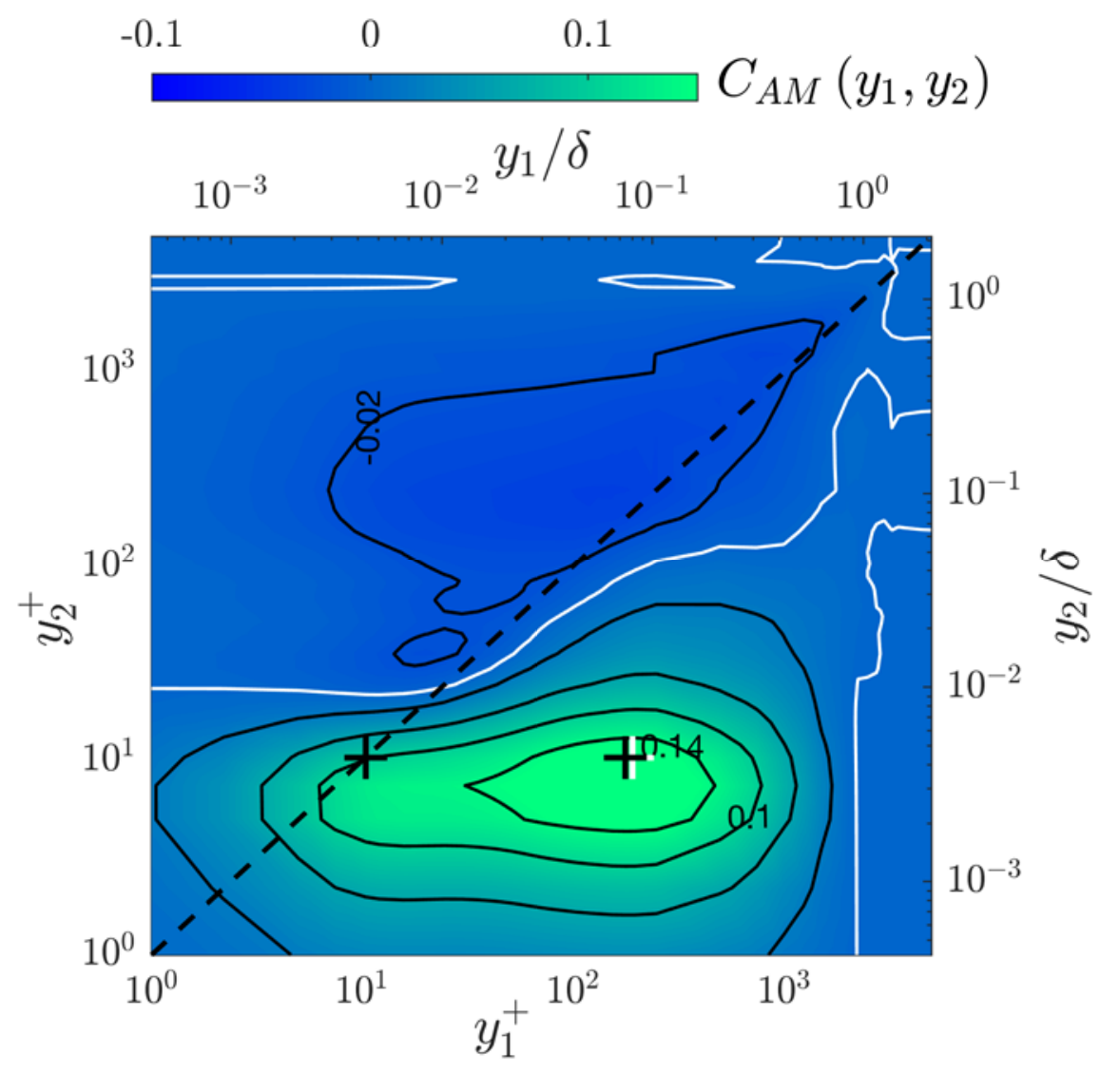} \label{fig:2p2DS_Hilbertspan_time}}
\subfigure[]{
\centering
\includegraphics[width=0.26\linewidth]{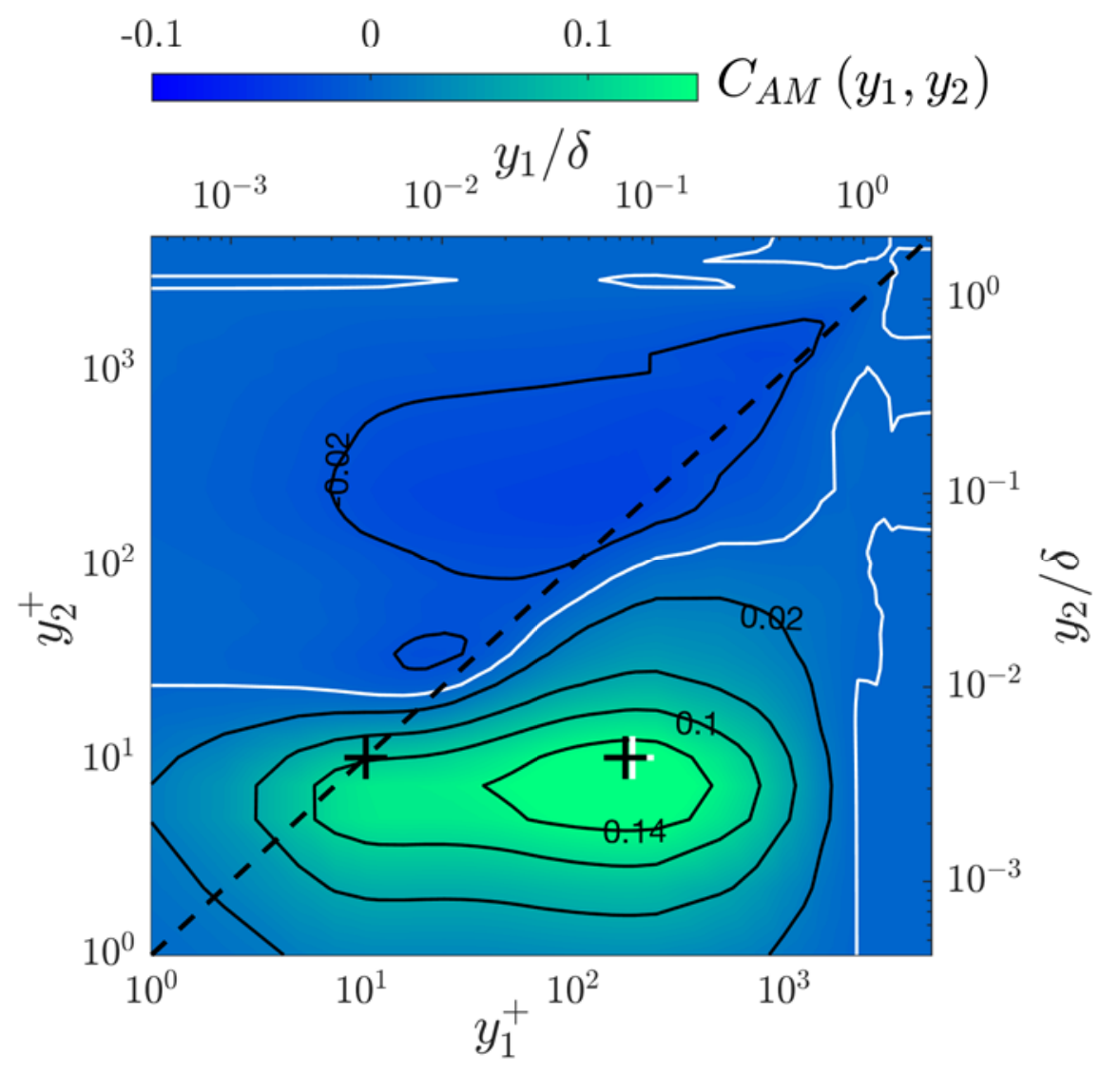} \label{fig:2p2DStime}}
\subfigure[]{
\centering
\includegraphics[width=0.26\linewidth]{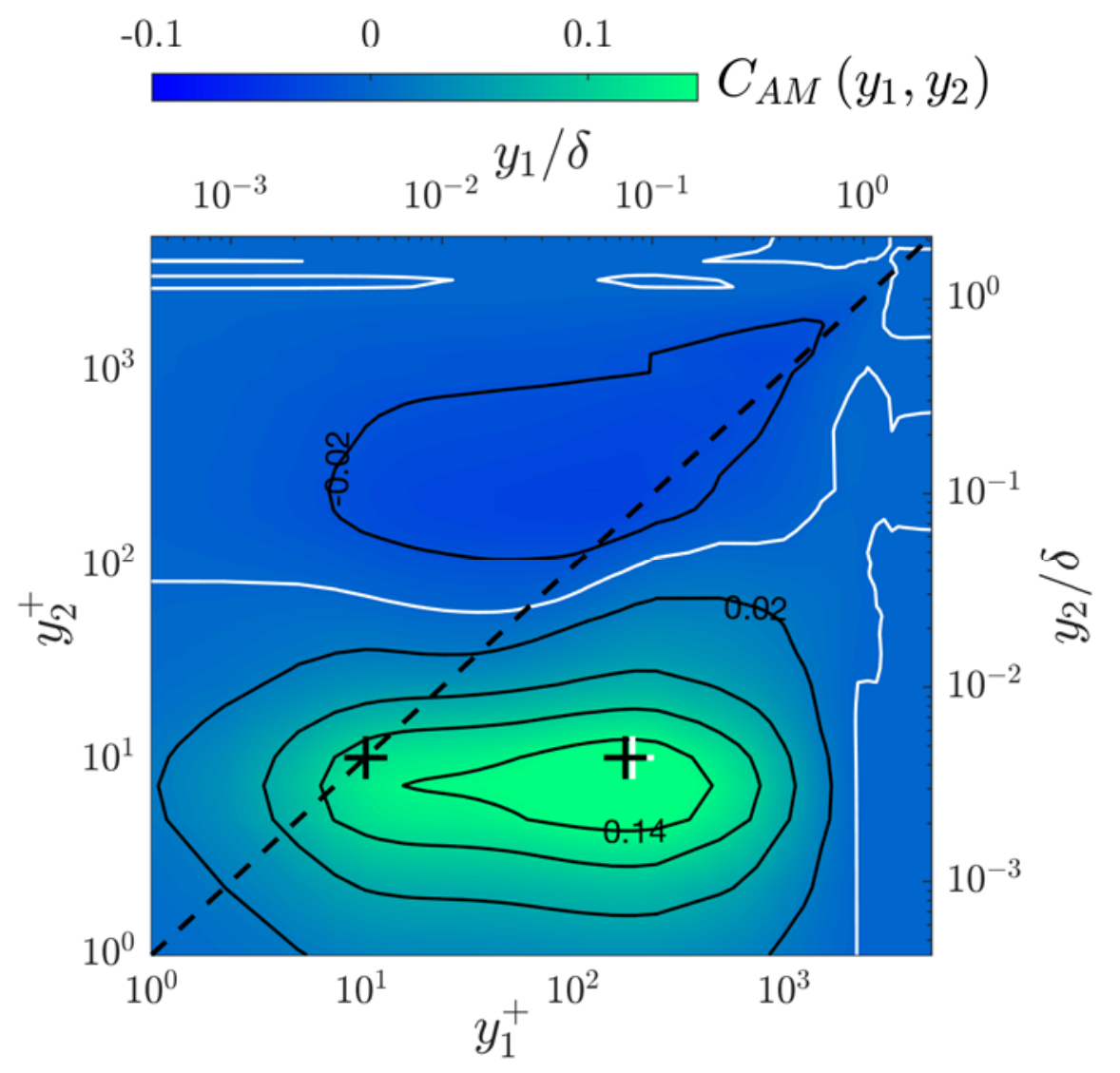} \label{fig:2p2DS}}
\subfigure[]{
\centering
\includegraphics[width=0.26\linewidth]{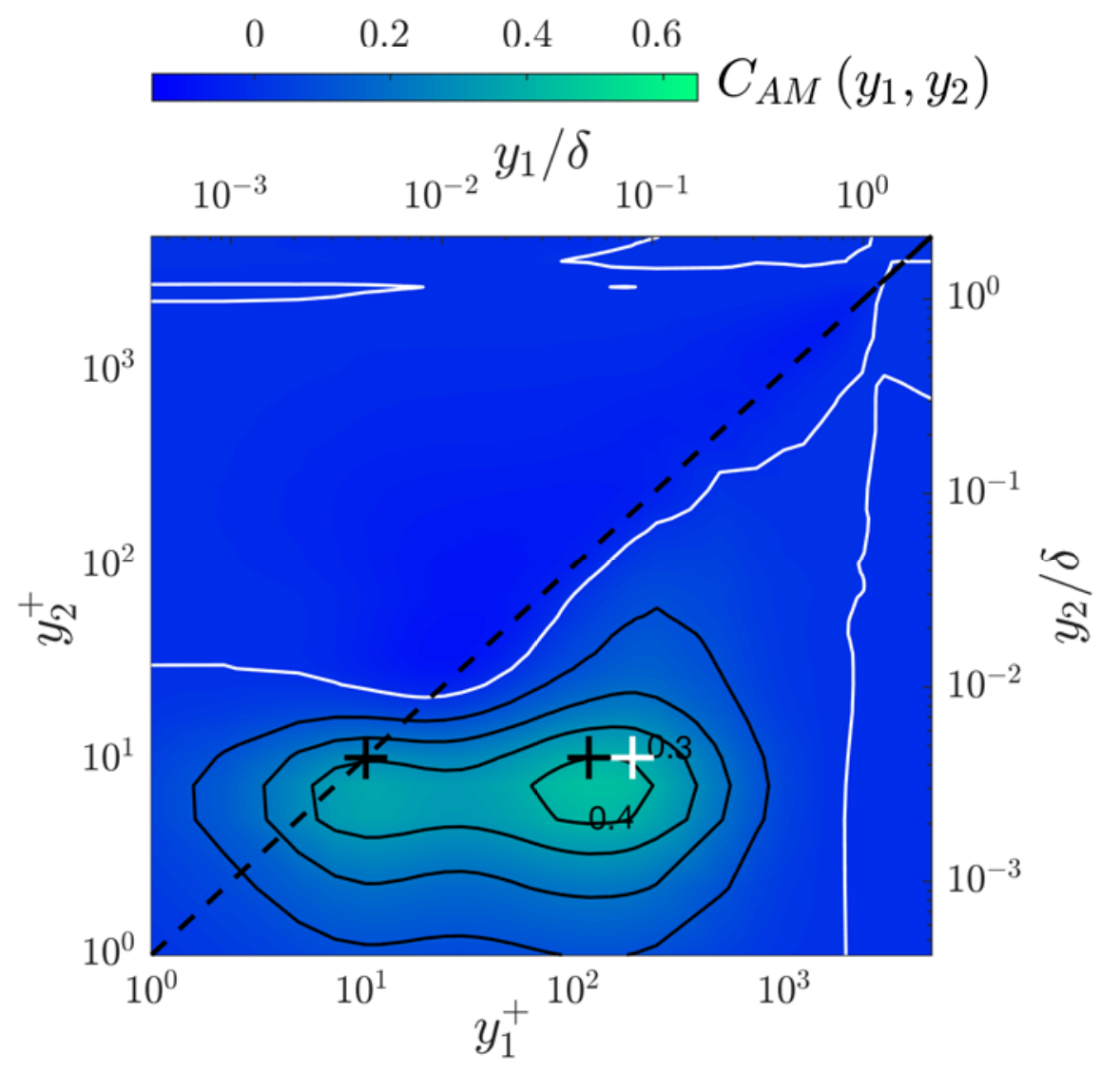} \label{fig:2pEMDtime}}
\vspace{-2mm}
\caption{\label{fig:cov2p} Contour maps of the covariance definition, 
$C_{AM}$, for $Re_\theta=8200$. White contour lines mark zero values. Some contour levels are shown for reference. Black crosses mark the locations $\left(y_1, y_2\right)$ for the covariance peaks in inner and outer region. White cross indicates the location for $\left(3.9Re_\tau^{0.5}, 10\right)$. Refer to Table \ref{tab:caption} for the details of each panel.}
\end{figure*}

Figure \ref{fig:cov2p} shows the contour maps of the covariance $C_{AM}$ for different filters, as detailed in Table \ref{tab:caption}.
Before comparing different panels, it is helpful to highlight the notable features of these contour maps. These maps typically depict two peaks, namely a diagonal and an off-diagonal peak. The formation of the off-diagonal peak is suggested to be a more refined picture of AM (Bernardini and Pirozzoli, 2011, Eitel-Amor \emph{et al.}, 2014). Bernardini and Pirozzoli (2011) observed that the off-diagonal peak seemed to disappear for a synthetic signal whereas the diagonal peak, {\it i.e.}\ the single-point correlation ($y_1$=$y_2$), was still present. This is consistent with the findings by Schlatter and \"Orl\"u (2010), and thus Bernardini and Pirozzoli concluded that the diagonal peak might have artifacts while the off-diagonal one conveys realistic information for the AM. For their data set, only spatial information was available; therefore every step of the calculation was performed in the spanwise direction only.
Here, Figure \ref{fig:2pspan} follows their approach and similarly depicts two peaks. However, when the filters are changed at different steps of the calculation compared to their case, the shape of the covariance maps does  change significantly. Figure \ref{fig:2pspantime} uses a time filter for the secondary filter and the distinct diagonal peak seems to disappear and this brings up the potential dependence of the diagonal peak on the secondary filtering of the envelope signal. However, this may not be an effect entirely on its own. The primary filtering to decompose the scales has also an impact on this effect, since similar results to  those in Figure \ref{fig:2pspantime} are obtained in Figure \ref{fig:2p2DS_Hilbertspan_time} when the 2DS filter is used for the primary filtering to decompose the scales. On the other hand, among the three, Figure \ref{fig:2p2DS_Hilbertspan_time}, \ref{fig:2p2DStime} and  \ref{fig:2p2DS}, there are no negligible differences, hinting on the more dominant effect of the primary filtering of the scale decomposition. In this regard, Figure \ref{fig:2pEMDtime} gives an informative result, when compared to \ref{fig:2p2DStime}: The correlation of the large scales with the small scales seems to extend in a larger area throughout the boundary layer with the 2DS Fourier filter than with EMD. Also, it is noticeable that the off-diagonal peak for EMD is not as strong in amplitude as for the 2DS Fourier filters. The assignment of the number of modes to each scale in EMD could affect the dominance of the large scales over the small scales. It is also worth noting here the potential limitation of the method as practised in the literature, that is to fix the number of modes for every wall-normal location. This might affect the correct representation of {\it e.g.}\ the increasing energy of the large scales in the outer region. 
\begin{table*}[htbp!]
\vspace{-2mm}
\small
\centering
\caption{\label{tab:caption} Details of the panels in Figure \ref{fig:cov2p}}
\label{tab}
\begin{tabular}{llll}
  \multicolumn{1}{c}{\textbf{Figure panel}} & \multicolumn{1}{c}{\textbf{Scale decomposition}} & \multicolumn{1}{c}{\textbf{Envelope calculation}} & \multicolumn{1}{c}{\textbf{Filtered envelope calculation}}   \\
a & spanwise filter & Hilbert in spanwise & spanwise filter \\
b & spanwise filter & Hilbert in spanwise & time filter \\
c & 2DS filter & Hilbert in spanwise & time filter   \\
d & 2DS filter & Hilbert in time & time filter   \\
e & 2DS filter & Hilbert in time & 2DS filter  \\
f & EMD, nSS = 5 & Hilbert in time & time filter \\
\end{tabular}
\normalsize
\end{table*}
%
%
%
\vspace{-2mm}
\subsection{Robustness}
\begin{figure*}[htbp!]
	\begin{center}
        \subfigure[]{\centering\includegraphics[scale=0.75]{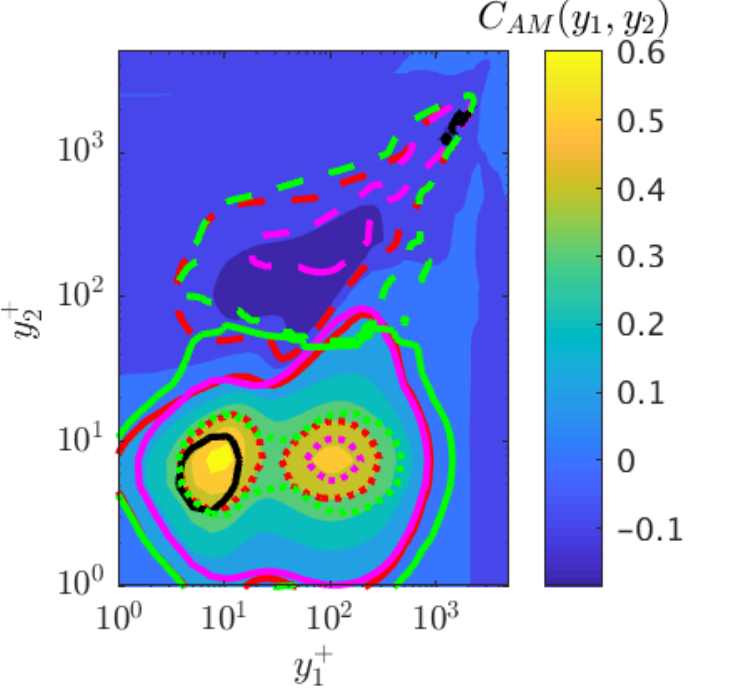}}   
		\subfigure[]{\centering\includegraphics[scale=0.75]{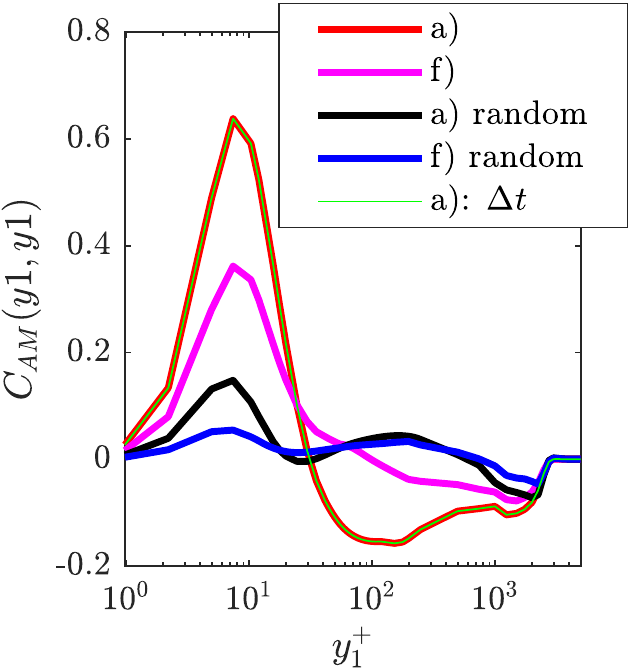}}\hspace{5pt}%
        \vspace{-2mm}
        \caption{Covariance $C_{AM}=\overline{u_L^+(y_1^+)E_L(u_S^+ (y_2^+))}$ computed for turbulent velocity signals at $Re_\theta=8200$, before and after having randomised the order of the data in the spanwise and (for case f) time direction. (a) Full map of the two-point covariance. Dashed, solid and dotted lines indicate, respectively, the contour levels $\left\{-0.05; 0.1; 0.4 \right\}$. (b) Single-point $y_1^+=y_2^+$ covariance. The background colour refers to $C_{AM}$ obtained with the spanwise filter (case a). The green line indicates the value of $C_{AM}$ obtained by shifting the $u_S^+$ signal in time until $C_{AM}$ is maximised.  
        \label{fig:CAM}}
    \end{center}
\end{figure*}

A sound measure of AM must be robust against the particular filtering procedure of choice, and not allow false positive identification of AM. In other words, it should predict modulation only when it is actually present in the signal, and return a vanishing measure in case the signal is constructed in a way that no modulation is present. In the following, these two aspects are addressed for two different measures of AM: the covariance defined previously, $C_{AM}$, and the simplified measure coming from the decomposed skewness, proposed by Eitel-Amor {\it et al.} (2014):
\begin{equation}
	C_{AM}^\ast=\overline{u_L^+(y_1^+) {u_S^+}^2 (y_2^+)}\, .
\end{equation}
Figure \ref{fig:CAM} shows $C_{AM}$ computed from the same turbulent boundary layer data at $Re_\theta=8200$ treated in different ways. In the baseline cases, the natural turbulent data has been fed directly to the primary and secondary filters required for decomposing into $u_S$ and $u_L$, and then for computing $E_L(u_S)$ (corresponding to cases a and f in Table \ref{tab}). In order to test the robustness against false positives, the inherent scale information is intentionally removed by randomly scrambling the signal along the directions across which the primary and secondary filter operate.  The spanwise Fourier filter and EMD (denoted a and f in Table \ref{tab}, respectively) are considered for both primary and secondary filtering, thus scrambling occurs along spanwise direction or along both spanwise direction and time, respectively. The randomised signal preserves all statistical moments of the original turbulent signal but contains no scale information or spatial coherence along the specified direction. In this way, no data values are generated, only their position is randomly changed. Thus, it features a flat spectrum typical of white noise. The randomisation occurs homogeneously in the wall-normal direction, in order to maintain the wall-normal coherence and covariance $\overline{u(y_1) u(y_2)}$ of the original turbulent signal prior to scale decomposition. Having done otherwise would have necessarily implied zero values of any two-point statistics at two different wall-normal heights, as to be expected from two independent random signals. 

\begin{figure*}[htbp!]
	\begin{center}
        \subfigure[]{\centering\includegraphics[scale=0.75]{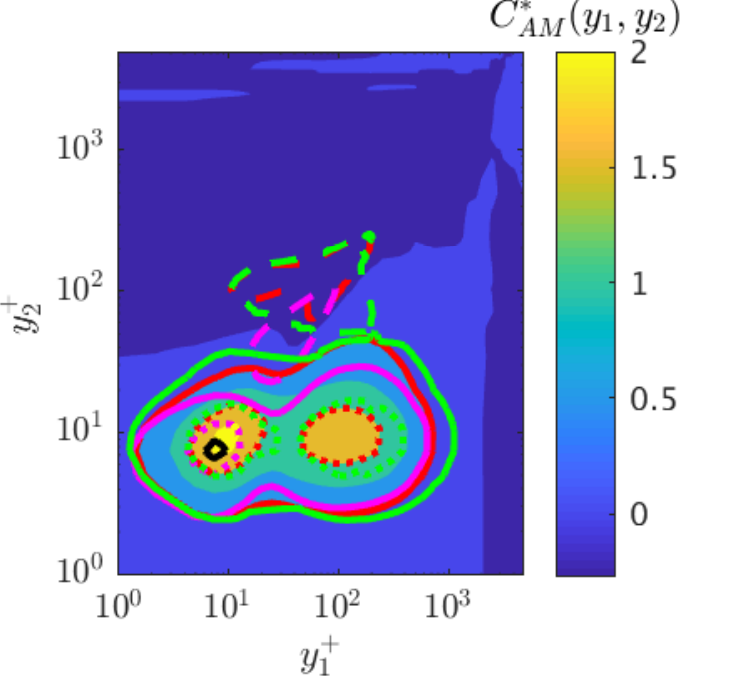}}
        \subfigure[]{\centering\includegraphics[scale=0.75]{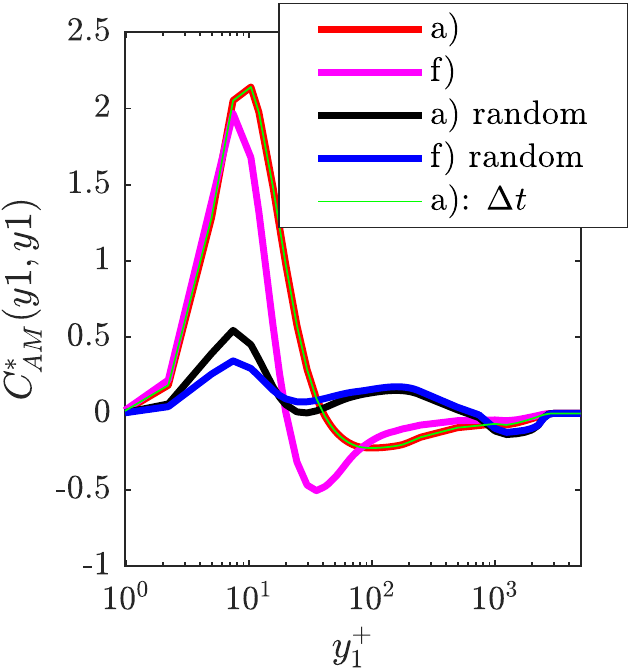}}\hspace{5pt}%
        \vspace{-2mm}
        \caption{Covariance $C_{AM}^\ast=\overline{u_L^+(y_1^+) {u_S^+}^2 (y_2^+)}$ computed for turbulent velocity signals at $Re_\theta=8200$. Colors as in Figure \ref{fig:CAM}. Dashed, solid and dotted lines indicate respectively the contour levels $\left\{-0.2; 0.5; 1.5 \right\}$. 
        \label{fig:CAMstar}}
    \end{center}
\end{figure*}

Let us first consider the effect of the primary and secondary filter pair on $C_{AM}$ maps for natural turbulent signals. As apparent from the figure, the filter choice affects $C_{AM}$ both quantitatively and qualitatively. Whereas the peak for small $y_2$ is visible for both filters, the log region features a significantly more negative AM for the Fourier filter. On the other hand, for a randomised signal, the covariance drops for both filters, only leaving a small spurious inner peak for the Fourier signal.
The same analysis is presented in Figure\ \ref{fig:CAMstar} for $C_{AM}^{*}$. It is interesting to note that in this case, the behaviour of the two filters is much more similar, even quantitatively. This can potentially be explained by the fact that one considers directly one term of the decomposed skewness, and is thus less affected by the exact details of the correlation, as long as a proper separation of large and small scales can be accomplished. Similarly, the randomised signal, having the same statistical moments, shows nearly no AM throughout the $y_1, y_2$ plane.
%
\subsection{Time shift and convection}
\begin{figure}[htbp!]
	\begin{center}
		\includegraphics[scale=0.75]{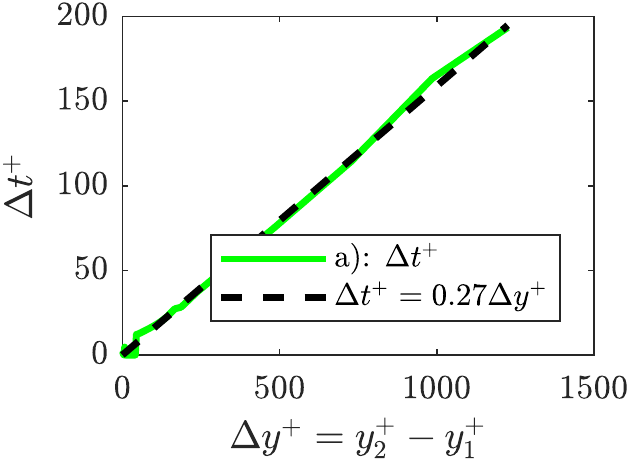}
        \vspace{-2mm}
        \caption{Time shift $\Delta t^+$ of the $u_S^+$ time series that maximises the covariance $C_{AM}^\ast=\overline{u_L^+(y_1^+,t) \left[ u_S^+ (y_2^+,t+\Delta t) \right]^2}$ at $y_1^+=10.3$  for the $Re_\theta=8200$ case. \label{fig:dt}}
    \end{center}
\end{figure}

As discussed in \emph{e.g.}\ Mathis {\it et al.}, due to the inclination of the large-scale structures in a boundary layer, the strongest AM is presumably not in the vertical direction, but rather along the inclination of structures; an angle of $\alpha=$10--15$^\circ$ is commonly reported in the literature. Using our LES data, we tested at each wall-normal position different time shifts between the small-scale and large-scale signals, and extracted a shift which maximises the correlation (or covariance). The corresponding results are shown in Figures\ \ref{fig:CAM}--\ref{fig:dt} (green lines). As noted by Bernardini and Pirozzoli (2011), there is only negligible influence of this time shift on the actual AM values, see Figures\ \ref{fig:CAM}(b) and \ref{fig:CAMstar}(b). This is most likely due to the fact that the structures are very long, and thus a small inclination will ultimately not change the quantification significantly. The actual time shift can be shown as a function of height for fixed $y_1^+=10.3$, see Figure\ \ref{fig:dt}. As expected, a larger height for the large-scale signal $y_2$ implies a larger time shift $\Delta t$. The plot indicates a nearly perfect linear fit, $\Delta t^+ = 0.27 \Delta y^+$. Assuming a convection velocity of around $0.8U_\infty$ for the large-scale structures, \emph{i.e.}\ $\Delta x^+ = 0.8{U_\infty^+} \Delta t^+$, one obtains $\Delta x^+ = 0.22{U_\infty^+} \Delta y^+$ (with ${U_\infty^+}\approx{27.5}$) which is close to the expected angle $\tan(\alpha)=\Delta y^+/ \Delta x^+$. One can thus conclude that it is indeed these large structures that are responsible for the AM. 
%
\section{ Conclusions }
We use a well-resolved LES data set by Eitel-Amor {\it et al.}\ (2014) to investigate the robustness of the amplitude modulation phenomenon, relevant for wall-bounded turbulence to quantify scale interactions. Both Fourier filters and EMD were tested to decompose the scales and both were found  effective. For quantifying, the definition by Mathis \emph{et al.}\ (2009) was adapted for two-point correlation maps, which are regarded to better represent AM. This definition requires two subsequent filterings, and the effects of different filters at each step were investigated and both affected the resulting maps both qualitatively and quantitatively. The primary filter in the correlation definition, \textit{i.e.}\ to decompose the scales, manifested a more dominant effect. Also, a randomised signal was used to check the robustness of AM definition against a particular filtering procedure. Finally, the inclination of the large-scale structures in the boundary layer was taken into account and negligible differences in AM quantification were observed. While comparing and contrasting with the existing literature on this topic, the current work provides the various analyses on a single data set allowing a robustness study and reveals important points for further amplitude modulation investigations.      
%
\Acknowledgments
Financial support provided by the Knut and Alice Wallenberg Foundation is gratefully acknowledged. DG is supported by the Priority Programme SPP 1881 Turbulent Superstructures of the DFG.

%
\begin{References}
\vspace{-1mm}
\item Agostini L. and Leschziner M. (2014), On the influence of outer large-scale structures on near-wall turbulence in channel flow, {\it Phys. Fluids}, Vol. 26, 075107.
\item Bernardini M. and Pirozzoli S. (2011), Inner/outer layer interactions in turbulent boundary layers: A refined measure for the large-scale amplitude modulation mechanism, {\it Phys. Fluids}, Vol. 23, 061701.
\item Bhuiyan S. M. A., Adhami R. R. and Khan J. F. (2008), Fast and Adaptive Bidimensional Empirical Mode Decomposition Using Order-Statistics Filter Based Envelope Estimation, {\it EURASIP J. Adv.}, 728356.
\item Cormier M., Gatti D. and Frohnapfel B. (2016), Interaction between inner and outer layer in drag-reduced turbulent flows, {\it Proc. Appl. Math. Mech.}, Vol. 16, 633-634.
\item Dogan E., {\"O}rl{\"u} R., Gatti D., Vinuesa R. and Schlatter P. (2018), Quantification of amplitude modulation in wall-bounded turbulence, {\it Fluid Dyn. Res.} (in press).
\item Eitel-Amor G., {\"O}rl{\"u} R. and Schlatter P. (2014), Simulation and validation of a spatially evolving turbulent boundary layer up to {R}e$_{\theta}$=8300, {\it Int. J. Heat Fluid Flow}, Vol. 47, pp. 57-69.
\item Hutchins N. and Marusic I. (2007), Large-scale influences in near-wall turbulence, {\it Phil. Trans. R. Soc. A}, Vol. 365, pp. 647-664.
\item Mathis R., Hutchins N. and Marusic I. (2009), Large-scale amplitude modulation of the small-scale structures in turbulent boundary layers, {\it J. Fluid Mech.}, Vol. 628, pp. 311-337.
\item Schlatter P., {\"O}rl{\"u} R. (2012), Turbulent boundary layers at moderate Reynolds numbers: inflow length and tripping effects, {\it J. Fluid Mech.}, Vol. 710, pp. 5-34.
\item Schlatter P., {\"O}rl{\"u} R. (2010), Quantifying the interaction between large and small scales in wall-bounded turbulent flows: a note of caution, {\it Phys. Fluids}, Vol. 22, pp. 051704.
\end{References}
\end{document}